\documentclass[12pt]{article}

\usepackage{epsfig}
\usepackage{amsmath}
\usepackage{graphicx}
\usepackage{cite}
\newcommand{\ps}{p\hspace{-0.44em}/\hspace{0.06em}}
\def\Title#1{\begin{center} {\Large {\bf #1} } \end{center}}
\usepackage{fancyhdr}
\newcommand{\eq}[1]{Eq.~(\ref{#1})}
\newcommand{\bb}{\ensuremath{B\!-\!\Bbar{}\,}}
\newcommand{\real}{\mathrm{Re}\,}

\begin{document}
\topskip 2cm 
\newcommand{\fig}[1]{Fig.~\ref{#1}} 
\newcommand{\bbms}{\bbs\ mixing}
\newcommand{\bbmd}{\bbd\ mixing}
\newcommand{\bbmq}{\bbq\ mixing}
\newcommand{\bbm}{\bb\ mixing}
\newcommand{\bbd}{\ensuremath{B_d\!-\!\Bbar{}_d\,}}
\newcommand{\bbs}{\ensuremath{B_s\!-\!\Bbar{}_s\,}}
\newcommand{\bbq}{\ensuremath{B_q\!-\!\Bbar{}_q\,}}
\newcommand{\Bbar}{\,\overline{\!B}}
\newcommand{\dd}{\ensuremath{D\!-\!\Dbar{}\,}}
\newcommand{\kk}{\ensuremath{K\!-\!\Kbar{}\,}}
\newcommand{\ddm}{\dd\ mixing}
\newcommand{\kkm}{\kk\ mixing}
\newcommand{\Dbar}{\,\overline{\!D}}
\newcommand{\Kbar}{\,\overline{\!K}}
\bibliographystyle{unsrt}

\Title{Flavor violation in the MSSM and
implications for top and squark searches at colliders}
\bigskip

\begin{raggedright}  

{\it \underline{Andreas Crivellin}  \index{}
\footnote{Presented at Linear Collider 2011: Understanding QCD at Linear Colliders  in searching for old and new physics, 12-16 September 2011, ECT*, Trento, Italy}\\
ITP, University of Bern, Switzerland\\
{\rm  crivellin@itp.unibe.ch}
}\\

\bigskip\bigskip
\end{raggedright}
\vskip 0.5  cm
\begin{raggedright}
{\bf Abstract} In this article I review some connections between flavor physics and collider physics. The first part discusses the effect of right-handed charged currents on the determination of the CKM elements $V_{ub}$ based on Ref.~\cite{Crivellin:2009sd}. It is shown that such an effective right-handed W-coupling can be generated in the MSSM which would lead to a sizable enhancement of single-top production at the LHC. The second part of this article focuses on the constraints on the mass splitting between left-handed squarks from Kaon and D mixing based on Ref.~\cite{Crivellin:2010ys}. Such a mass splitting has interesting consequences for squark decay chains at colliders.

\end{raggedright}

\section{Right-handed W-coupling}

In the standard model (SM) the tree-level W coupling has a pure $V-A$
structure meaning that all charged currents are left-handed.
Right-handed charged currents were first studied in the context of
left-right symmetric models \cite{Senjanovic:1975rk} which enlarge the
gauge group by an additional $SU(2)_R$ symmetry between right-handed
doublets. In these models new right-handed gauge bosons $W_R$, $Z_R$
appear and the physical SM-like W-boson has a dominant left-handed
component with a small admixture of $W_R$. The latter will generically
lead to small right-handed couplings to both quarks and leptons.  The
right-handed mass scale inferred from today's knowledge on neutrino
masses is so large that all right-handed gauge couplings are
undetectable. Most of these couplings are further experimentally
strongly constrained \cite{Amsler:2008zzb}.  A different source of
right-handed couplings of quarks to the W-boson can be loop effects,
which generate a dimension-6 quark-quark-W vertex.  In this case no
right-handed lepton couplings occur, as long as the neutrinos are
assumed left-handed.

\subsection{Right-handed W couplings}
An appropriate framework for our analysis is an effective Lagrangian. 
Following the notation of Ref.~\cite{Grzadkowski:2008mf}, we write
\begin{equation}
{\cal L} = {\cal L}_{\rm SM} 
+ \frac{1}{\Lambda  } \sum_i C_i^{(5)} Q_i^{(5)} 
+ \frac{1}{\Lambda^2} \sum_i C_i^{(6)} Q_i^{(6)} 
+ {\cal O}\left(\frac{1}{\Lambda^3}\right),
\label{Leff}
\end{equation}
here ${\cal L}_{\rm SM}$ is the standard model (SM) Lagrangian, while
$Q_i^{(n)}$ stand for dimension-$n$ operators built out of the SM fields
and beeing invariant under the SM gauge symmetries. Such an effective theory approach is
appropriate for any SM extension in which all new particles are
sufficiently heavy ($M_{\rm new} \sim \Lambda \gg m_t$). As long as only
processes with momentum scales $\mu \ll \Lambda$ are considered, all heavy
degrees of freedom can be eliminated \cite{Appelquist:1974tg}, leading to the
effective theory defined in (\ref{Leff}). The operators $Q_i^{(5)}$ and $Q_i^{(6)}$ have been completely classified in Ref.~\cite{Buchmuller:1985jz}. Here, we need the following dimension-six operator describing anomalous (not present in the SM) right-handed W-couplings to quarks:
\begin{equation}
  Q_{RR}=\bar{u}_f \gamma^\mu P_R d_i 
  \left(\tilde \phi ^\dagger i D_\mu \phi\right)+h.c.
\label{Operator}
\end{equation}
where $\phi$ denotes the Higgs doublet $D_\mu$ is the covatrinat derivative and $\widetilde{\phi} =
i\tau^2\phi^*$. The Feynman rule for the $W$-$u_f$-$d_i$ interaction vertex,
\begin{equation}
\frac{ - i g_2 \gamma^\mu}{\sqrt 2 }
 \left( {V_{fi}^L P_L  + V_{fi}^R P_R } \right)
, \label{W-coupling}
\end{equation}
is found by combining the usual SM interaction with the extra
contributions that are obtained by setting the Higgs field in
\eq{Operator} to its vacuum expectation value.  In \eq{W-coupling}
$V^L_{fi}$ and $V^R_{fi}$ denote elements of the effective CKM matrices, which are not
necessarily unitary.  $V_{fi}^R$ is related to the Wilson coefficient in
\eq{Leff} via $V_{fi}^R = \frac{C_{RR}}{2 \sqrt{2} G_F \Lambda^2}$.
$V_{fi}^L$ receives contributions from the tree-level CKM matrix and 
the LL analogue of $Q_{RR}$ in \eq{Operator}. 

In Ref.~\cite{Grzadkowski:2008mf} it was pointed out that very strong
constraints can be obtained on $V_{tb}^R$ from $b\to s\gamma$, because
the usual helicity suppression factor of $m_b/M_W$ is absent in the
right-handed contribution.  By the same argument $V_{td}^R$ is tightly constrained.
Large effects concerning transitions between the first two generations
are unlikely, because $V_{us,cd}^L$ are larger than other
off-diagonal CKM elements. Thus, we focus our attention 
on the remaining element $V_{ub}^R$ (similar effects are possible for $V_{cb}^R$ but the signature is less significant).

\begin{figure}
\centering
\includegraphics[width=0.47\textwidth]{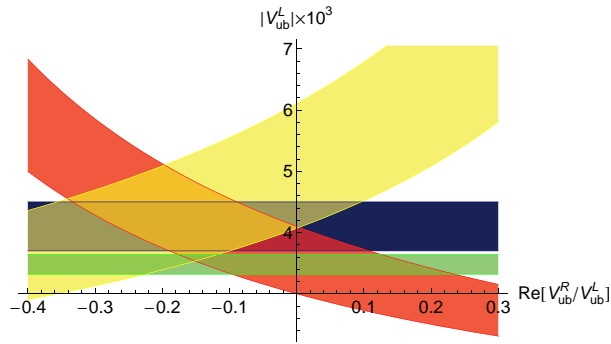}
\caption{\label{Vub} \small $\left|V^L_{ub}\right|$ as a function of
  $\rm{Re}\left[V^R_{ub}/V^L_{ub}\right]$ extracted from different
  processes. Blue(darkest): inclusive decays. Red(gray): $B\to \pi l\nu$.
  Yellow(lightest gray): $B\to \tau\nu$. Green(light gray): $V^L_{ub}$ determined from CKM unitarity.}
  \vspace{-15mm}
\end{figure}

The experimental determination of $|V_{ub}|$ from both
inclusive and exclusive $B$ decays is a mature field by now
\cite{Amsler:2008zzb}. To discuss the
impact of right-handed currents we denote the CKM element extracted from
data with SM formula by $V_{ub}$. If the matrix
element of a considered exclusive process is proportional to the vector
current, $V_{ub}^L$ and $V_{ub}^R$ enter with the same sign and the
´´true'' value of $V_{ub}^L$ in the presence of $V_{ub}^R$ is given by:
\begin{equation}
V_{ub}^L=V_{ub}-V_{ub}^R \label{truel}
\end{equation}
For processes proportional to the axial-vector current $V_{ub}^R$ enters
with the opposite sign as $V_{ub}^L$, so that
\begin{equation}
V_{ub}^L=V_{ub}+V_{ub}^R . \label{truer}
\end{equation}

In inclusive decays the interference term between the left-handed and
right-handed contributions is suppressed by a factor of $m_u/m_b$ so
that it is irrelevant for $V_{ub}$. The remaining dependence on $V_{ub}^R$ is
quadratic and therefore negligible. 
Note that the determinations from inclusive
and exclusive semileptonic decays agree within their errors, but the
agreement is not perfect \cite{Amsler:2008zzb,Charles:2004jd}. The
analysis of $B\to\tau\nu$ is affected by the uncertainty in the
decay constant $f_B$. Within errors the three determinations of
$|V_{ub}|$ are compatible for $V_{ub}^R=0$, as one can read off from \fig{Vub}.  The
picture looks very different once the information from a global fit to
the unitarity triangle (UT) is included: As pointed out first by the
CKMFitter group, the measured value of $B\to\tau\nu$ suffers from a
tension with the SM of 2.4--2.7$\sigma$ \cite{Charles:2004jd}.  First,
the global UT fit gives a much smaller error on $|V_{ub}|$ (as a
consequence of the well-measured UT angle $\beta$); the corresponding
value is also shown in \fig{Vub}.  Second, the data on \bbmd\ exclude
very large values for $f_B$, which in turn cuts out the lower part of the
yellow (light gray) region in \fig{Vub}. Essentially we realize from
\fig{Vub} that we can remove this tension while simultaneously bringing
the determinations of $|V_{ub}|$ from inclusive and exclusive
semileptonic decays into even better agreement. For this the
right-handed component must be around $\real (V_{ub}^R/V_{ub}^L) \approx
-0.15$. 

\subsection{MSSM renormalization of the quark-quark-W vertex}

\begin{figure}
\centering
\includegraphics[width=0.4\textwidth]{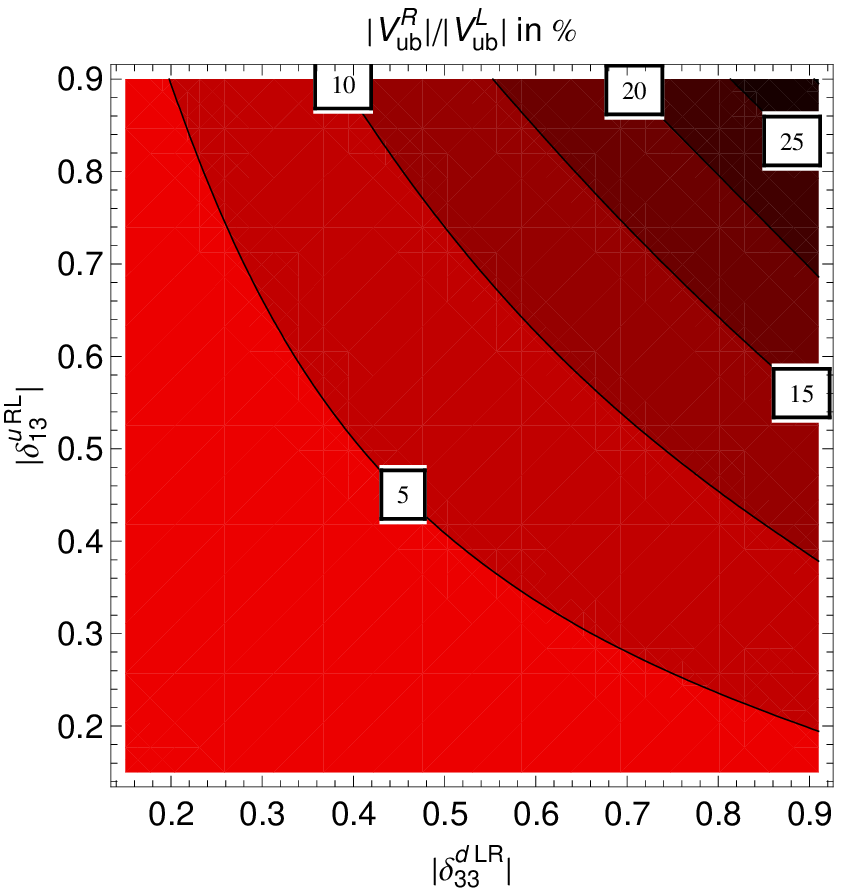}
\includegraphics[width=0.55\textwidth]{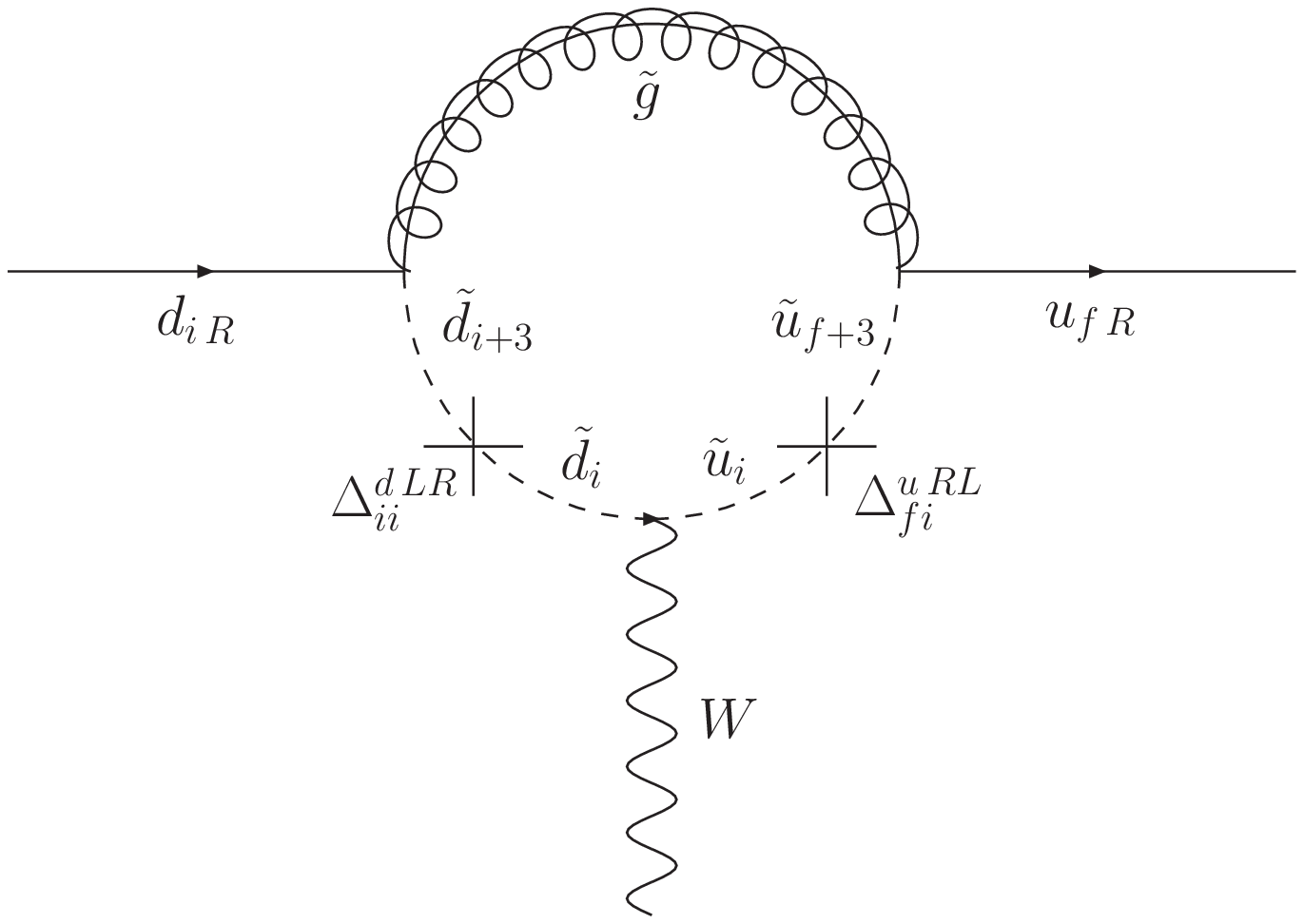}
\caption{\label{rvubcb} \small
 Right: Feynman diagram which induces the effective right-handed W coupling of a down-type quark of flavor i to an up-type quark of flavor f. The crosses stand for the flavor and chirality changes needed to generate the coupling.\newline
Left: Relative strength of the induced right-handed
  coupling $|V^R_{ub}|$ with respect to $|V^L_{ub}|$
  for $M_{\rm{SUSY}}=1\,\rm{TeV}$. $|V^L_{ub}|$ is determined from CKM unitarity.}
    \vspace{-15mm}
  \end{figure}

In Ref.~\cite{Crivellin:2008mq} the renormalization of the quark-quark-W
vertex by non-decoupling chirally enhanced supersymmetric self-energies has been
computed. Here, we extend this analysis and calculate the leading contributions to the 
quark-quark-W vertex which decouple for $M_{\rm SUSY}\to \infty$.   
Using the conventions of Ref.~\cite{Crivellin:2008mq} we expand to
first oder in the external momenta and decompose the self-energies as
\begin{eqnarray}
  \Sigma _{fi}^q  &=& \;\;\;\;\left( {\Sigma _{fi}^{q\;LR}  + \ps\Sigma
      _{fi}^{q\;RR} } \right)P_R \nonumber \\
&&\,+ \left( {\Sigma _{fi}^{q\;RL}  +
      \ps\Sigma _{fi}^{q\;LL} } \right)P_L . 
\end{eqnarray}
These self-energies lead to a flavor-valued wave-function
renormalization $\Delta U_{fi}^{q\;L,R}$ for all external left- and
right-handed fields. It is useful to decompose these factors further in
to an unphysical anti-Hermitian part $\Delta U_{fi}^{q\;L\;A}$, which can
be absorbed into the renormalization of the CKM matrix, and a Hermitian
part $\Delta U_{fi}^{q\;L\;H}$, which can constitute a physical effect 
appearing as a deviation from CKM unitarity: $\Delta U_{fi}^{q\;L,R\;H}  = \Sigma _{fi}^{q\;LL,RR}/2$. 
Neglecting external momenta, the genuine vertex-correction originating
from a squark-gluino loop is given by
\begin{align}
 &- i\Lambda _{u_f d_i }^{W\;\tilde g}  =  \label{vertexkorrktur}\\
 & \dfrac{{g_2 }}{{\sqrt 2 }}\dfrac{{i\alpha _s }}{{3\pi }}\gamma ^\mu \!\! \sum\limits_{s,t = 1}^6 {\sum\limits_{j,k = 1}^3 \!{\left( {W_{fs}^{\tilde u} W_{ks}^{\tilde u*} V_{kj}^{L} W_{jt}^{\tilde d} W_{it}^{\tilde d *} P_L  + W_{f + 3,s}^{\tilde u} W_{ks}^{\tilde u *} V_{kj}^{L} W_{jt}^{\tilde d} W_{i + 3,t}^{\tilde d*} P_R } \right)} C_2 \left( {m_{\tilde u_s } ,m_{\tilde d_t } ,m_{\tilde g} } \right)} . \nonumber
\end{align}
The matrices $W_{st}^{\tilde q}$ diagonalize the squark mass matrices
\cite{Crivellin:2008mq}. The part proportional to $P_L$ in \eq{vertexkorrktur} cancels
with the anti-Hermitian part of the wave-function renormalization
due to the SU(2) relation between the left-handed up and down
squarks for $M_{\rm SUSY}\to \infty$ according to the
decoupling theorem \cite{Appelquist:1974tg}. Since the loop function $C_2$
depends only weakly on $M_{\rm SUSY}$, the cancellation is very
efficient, even for light squarks around 300$\,$GeV. Therefore, the unitarity of the CKM matrix is conserved with very high accuracy.
A right-handed coupling of quarks to the W boson is induced by the diagram in Fig.~\ref{rvubcb} if
left-right mixing of squarks is present. The effective coupling
corresponds to $Q_{RR}$ in \eq{Operator} and vanishes in the
decoupling limit $M_{\rm{SUSY}}\to \infty$. There is no wave-function renormalization of right-handed
quarks which can be applied to the W vertex, therefore no gauge cancellations occur.

We show the relative size of the right-handed coupling involving u,c and b in Fig.~\ref{rvubcb}.
Note that the mass insertion $\delta^{u\;RL}_{13}$ is not affected by the fine-tuning
argument imposed in \cite{Crivellin:2008mq} nor severely restricted by
FCNC processes. Therefore, the size of the induced couplings
$V^R_{ub}$ can be large enough to explain
(attenuate) the apparent discrepancies among the various determinations
of $|V_{ub}|$. 

\subsection{Right-handed W coupling and single-top production}

We have seen that the disturbing problem with $B\to\tau\nu_\tau$
\cite{Charles:2004jd} can be removed and the inclusive and
exclusive determinations of $|V_{ub}|$ can be brought into agreement.
If one wants to achieve this in the MSSM a large left-right mixing
between sbottoms (as present in e.g.\ the popular large-$\tan\beta$
scenarios) and a large $A^u_{31}$-term is needed. Large values for $A^u_{31}$ enhance single-top production, making it observable at the LHC. If $\delta^{u\;RL}_{13}\approx0.6$ a 95\% CL signal can already be detected with 50 inverse femtobarn \cite{Plehn:2009it}. 

\section{Non-degenerate squark masses}

Already in the early stages of minimal supersymmetric standard model (MSSM) analyses it was immediately noted, that a super GIM mechanism is needed in order to satisfy the bounds from flavor changing neutral currents (FCNCs) \cite{Dimopoulos:1981zb}. Therefore, the mass matrix of the left-handed squarks should be (at least approximately) proportional to the unit matrix, since otherwise flavor off-diagonal entries arise inevitably either in the up or in the down sector due to the SU(2) relation between the left-handed squark mass terms (i. e. left-handed up squark and down squark mass matrices differ only by a CKM rotation). 
The idea that nondegenerate squarks can still satisfy the FCNC constraints (K and D mixing) was first discussed in Ref.~\cite{Nir:1993mx} in the context of Abelian flavor symmetries.

The squark spectrum is also a hot topic concerning bench-mark scenarios for the LHC. It is commonly assumed that the squarks are degenerate at some high scale and that non-degeneracies are introduced via the renormalization group \cite{sps,gamma}. In such scenarios, the non-degeneracies are proportional to Yukawa couplings and therefore only sizable for the third generation. In principle, there remains the possibility that squarks have already different masses at some high scale. The question to be clarified is which regions in parameter space with non-degenerate squarks are compatible with \dd\ and \kk\ mixing. 

\begin{figure}
\includegraphics[width=0.5\textwidth]{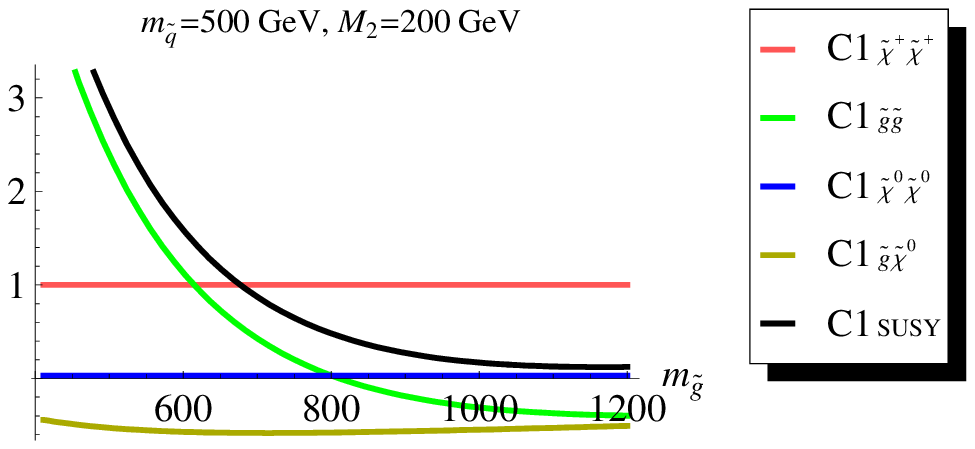}
\includegraphics[width=0.5\textwidth]{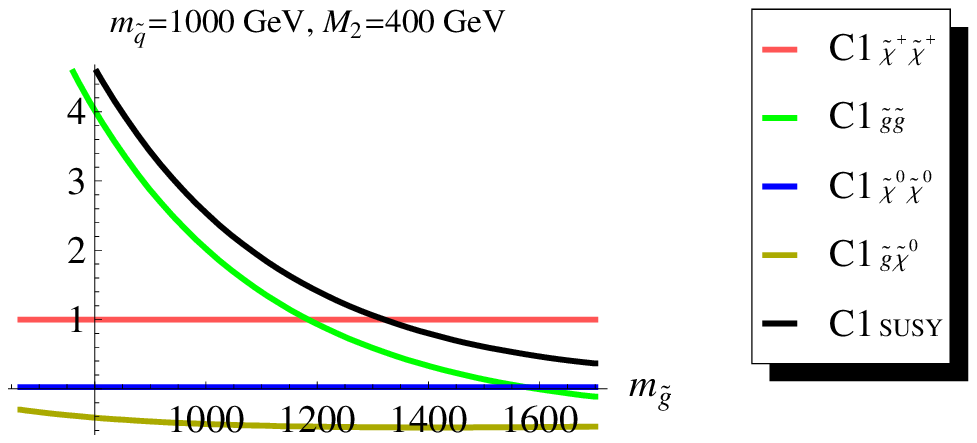}
\caption{\small Size of the real part of Wilson coefficients [see Eqs. (\ref{C1gg}) and (\ref{C1ew})] contributing to \dd\ or \kk\ mixing normalized to the chargino contribution as a function of $m_{\tilde{g}}$ for different values of $m_{\tilde{q}}$ and $M_2$ assuming a small nonzero (real) off-diagonal element $\delta^{q\;LL}_{12}$. $C_{1\rm{SUSY}}$ is the sum of all Wilson coefficients contributing in addition to the SM one. The relative size of the coefficients remains unchanged also in the case of complex elements $\delta^{q\;LL}_{12}$.}\label{C1}
  \vspace{-15mm}
\end{figure}

\subsection{Meson mixing between the first two generations}

Measurements of flavor-changing neutral current (FCNC) processes put strong constraints on new physics at the TeV scale and provide an important guide for model building. In particular \dd\ and \kk\ mixing strongly constrain transitions between the first two generations and combining both is especially powerful to place bounds on new physics \cite{Blum:2009sk}. 
In the down sector FCNCs between the first two generations are probed by the neutral Kaon system. Here the experimental values for the mass difference and the CP violating quantity $\epsilon_K$ are \cite{PDG}:
\begin{eqnarray}
\Delta m_K/m_K=(7.01\pm0.01)\times10^{-15}\nonumber\\
\epsilon_K=(2.23\pm0.01)\times10^{-3}
\label{Kmixing}
\end{eqnarray}
As we see from \eq{Kmixing} both the mass difference and the size of the indirect CP violation are tiny and the numbers are in agreement with the standard model (SM) prediction: The SM contribution to the mass difference is small due to a rather precise GIM suppression (the top contribution is suppressed by small CKM elements) and also the CP asymmetry is strongly suppressed because CP violation necessarily involves the tiny CKM combination $V_{td}V_{ts}^*$ related to the third fermion generation. Therefore, Kaon mixing puts very strong bounds on NP scenarios like the MSSM. According to the analysis of Ref.~\cite{Ciuchini:2000de} the allowed range in the $C_{M_K}-C_{\epsilon_K}$ plane is rather limited. At 95\% confidence level on can roughly expect the NP contribution to the mass difference $\Delta M_K$ to be at most of the order of the SM contribution. The NP contribution to $\epsilon_K$ is even more restricted. 

In the up sector FCNCs are probed by \dd\ mixing. In contrast to the well-established Kaon mixing, it was only discovered recently in 2007 by the BABAR and BELLE collaborations. The current experimental values are \cite{PDG}: 
\begin{eqnarray}
\Delta m_D/m_D=(8.6\pm2.1)\times10^{-15}\nonumber\\
A_\Gamma=(1.2\pm2.5)\times10^{-3}
\end{eqnarray}
Short-distance SM effects are strongly CKM suppressed and the long-distance contributions can only be estimated. Therefore, conservative estimates assume for the SM contribution a range up to the absolute measured value of the mass difference. However, due to the small measured mass difference D mixing still limits NP contributions in a stringent way. Furthermore, a CP violating phase in the neutral D system can directly be attributed to NP. 

In summary, \dd\ and \kk\ mixing restrict FCNC interactions between the first two generations in a stringent way and one should expect the NP contributions to the mass difference to be smaller than the experimental value \cite{Blum:2009sk}:
\begin{equation}
\Delta m^{\rm{NP}}_{D,K}\leq \Delta m^{\rm{exp}}_{D,K}
\label{DeltaM}
\end{equation}
CP violation associated with new physics is even more restricted, especially in the d sector:
\begin{equation}
\epsilon^{\rm{NP}}_{K}\leq 0.6\epsilon^{\rm{exp}}_{K}
\label{epsilon}
\end{equation}
Equations (\ref{DeltaM}) and (\ref{epsilon}) summarize in a concise way the allowed range for NP and we will use them to constrain the NP contributions to K and D mixing in Sec.~\ref{DK}.

\subsection{Constraints on the mass splitting from Kaon and D mixing.\label{DK}}

In the common definition of MFV \cite{Isidori} flavor-violation due to NP is postulated to stem solely from the Yukawa sector, resulting in FCNC transitions proportional to products of CKM elements and Yukawa couplings. Therefore, such scenarios allow only sizable deviations from degeneracy with respect to the third generation. A more general notion of MFV could be defined by stating that all flavor changes should be induced by CKM elements. This definition would also cover the case with a diagonal squark mass matrix in one sector (either the up or the down sector) but with off-diagonal elements, introduced by the $SU(2)$ relation, in the other sector. This setup corresponds to an exact alignment of the squark mass term $m_{\tilde{q}}^2$ with the product of Yukawa matrices $Y_u^{\dagger}Y_u$ (or with $Y_d^{\dagger}Y_d$ in the case of a diagonal down squark mass matrix).

The obvious way how off-diagonal elements of the squark mass matrices enter meson mixing is via squark-gluino diagrams. These contributions to $O_1=\bar{s}\gamma^{\mu}P_L d\otimes \bar{s}\gamma_{\mu}P_L d$ are commonly expected to be dominant since they involve the strong coupling constant:
\begin{equation}
C_1^{\tilde g\tilde g}  =  - \frac{{g_s^4 }}{{16\pi ^2 }}\sum\limits_{s,t = 1}^6 {\left[ {\frac{{11}}{{36}}D_2 \left( {m_{\tilde q_s }^2 ,m_{\tilde q_t }^2 ,m_{\tilde g}^2 ,m_{\tilde g}^2 } \right) + \frac{1}{9}m_{\tilde g}^2 D_0 \left( {m_{\tilde q_s }^2 ,m_{\tilde q_t }^2 ,m_{\tilde g}^2 ,m_{\tilde g}^2 } \right)} \right]V_{s\;12}^{q\;LL} V_{t\;12}^{q\;LL} } 
\label{C1gg}
\end{equation}
Our conventions for the loop-functions and the matrices in flavor space $V_{s\;12}^{q\;LL}$ are given in the appendix of Ref.~\cite{Crivellin:2008mq}. However, if we have flavor-changing LL elements it is no longer possible to concentrate on the gluino contributions for four reasons:
\begin{itemize}
\vspace{-2mm}
	\item The gluino contributions suffer from cancellations between the boxes with crossed and uncrossed gluino lines corresponding to the two terms in the square brackets in \eq{C1gg}. The crossed box diagrams occur since the gluino is a majorana particle. This cancellation occurs approximately in the region where $m_{\tilde{g}}\approx 1.5 \,m_{\tilde{q}}$. \vspace{-2mm}
	\item In the SU(2) limit with unbroken SUSY the winos couple directly to left-handed particles with the weak coupling constant $g_2$. Therefore, flavor-changing LL elements can contribute without involving small left-right or gaugino mixing angles.
	\vspace{-2mm}
	\item Since charginos are Dirac fermions, there are no cancellations between different diagrams at the one-loop order. 
	\vspace{-2mm}
	\item The wino mass $M_2$ is often assumed to be much lighter than the gluino mass. In most GUT models the relation $M_2\approx m_{\tilde{g}}\alpha_{2}/\alpha_{3}$ holds. Since the loop function is always dominated by the heaviest mass, one can expect large chargino and neutralino contributions if the squarks masses are similar to the lighter chargino masses.
	\vspace{-2mm}
\end{itemize}
Therefore, we have to take into account the weak (and the mixed weak-strong) contributions to $C_1$: 
\begin{equation}
\begin{array}{l}
C_1^{\tilde \chi ^0 \tilde \chi ^0 }  =  - \frac{1}{128\pi ^2 }\frac{g_2^4 }{4}\sum\limits_{s,t = 1}^6 \left( D_2 \left( m_{\tilde q_s }^2 ,m_{\tilde q_t }^2 ,M_2^2 ,M_2^2  \right) + 2 M_2^2 D_0 \left( m_{\tilde q_s }^2 ,m_{\tilde{q}_t }^2 ,M_2^2 ,M_2^2  \right) \right)V_{s\;12}^{q\;LL} V_{t\;12}^{q\;LL}  
\\
C_1^{\tilde g\tilde \chi ^0 }  =  - \frac{1}{16\pi ^2 }\frac{g_s^2 g_2^2 }{2}\sum\limits_{s,t = 1}^6 \left( \frac{1}{6}D_2 \left( m_{\tilde q_s }^2 ,m_{\tilde q_t }^2 ,m_{\tilde g}^2 ,M_2^2  \right) + \frac{1}{3}m_{\tilde g} M_2 D_0 \left( m_{\tilde q_s }^2 ,m_{\tilde q_t }^2 ,m_{\tilde g}^2 ,M_2^2  \right) \right)V_{s\;12}^{q\;LL} V_{t\;12}^{q\;LL} 
\\
C_1^{\tilde \chi ^ +  \tilde \chi ^ +  } =  - \frac{g_2^4}{128\pi ^2 }\sum\limits_{s,t = 1}^6 D_2 \left( m_{\tilde q_s }^2 ,m_{\tilde q_t }^2 ,M_2^2 ,M_2^2  \right)V_{s\;12}^{q\;LL} V_{t\;12}^{q\;LL} 
\end{array} \label{C1ew}
\end{equation}
In \eq{C1ew} we have set all Yukawa couplings to zero and neglected small chargino and neutralino mixing. Because of the small Yukawa couplings of the first two generations and the suppressed bino-wino mixing the only sizable contribution of both the gluino and the electroweak diagrams is to the same operator $O_1=\bar{s}\gamma^{\mu}P_L d\otimes \bar{s}\gamma_{\mu}P_L d$ as the SM contribution. Note that in all contribution the same combination of mixing matrices enters, since the CKM matrices in the chargino vertex cancels with the ones in the squark mass matrix. 

In \fig{C1} we show the size of the different contributions to $C_1$ as a function of the gluino mass. We have normalized all coefficients to $C_1^{\tilde \chi ^ +  \tilde \chi ^ + }$ since only one box diagram contributes to it and therefore the coefficient depends only on one loop-function which is strictly negative. 

\begin{figure}[t]
\centering
\includegraphics[width=0.35\textwidth]{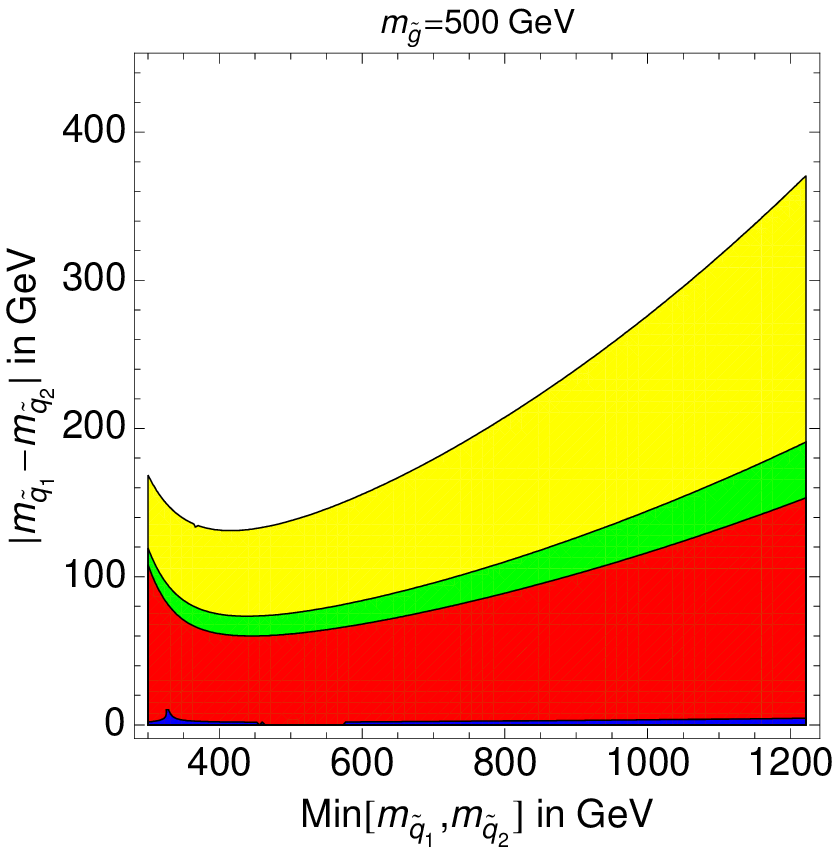}
\includegraphics[width=0.35\textwidth]{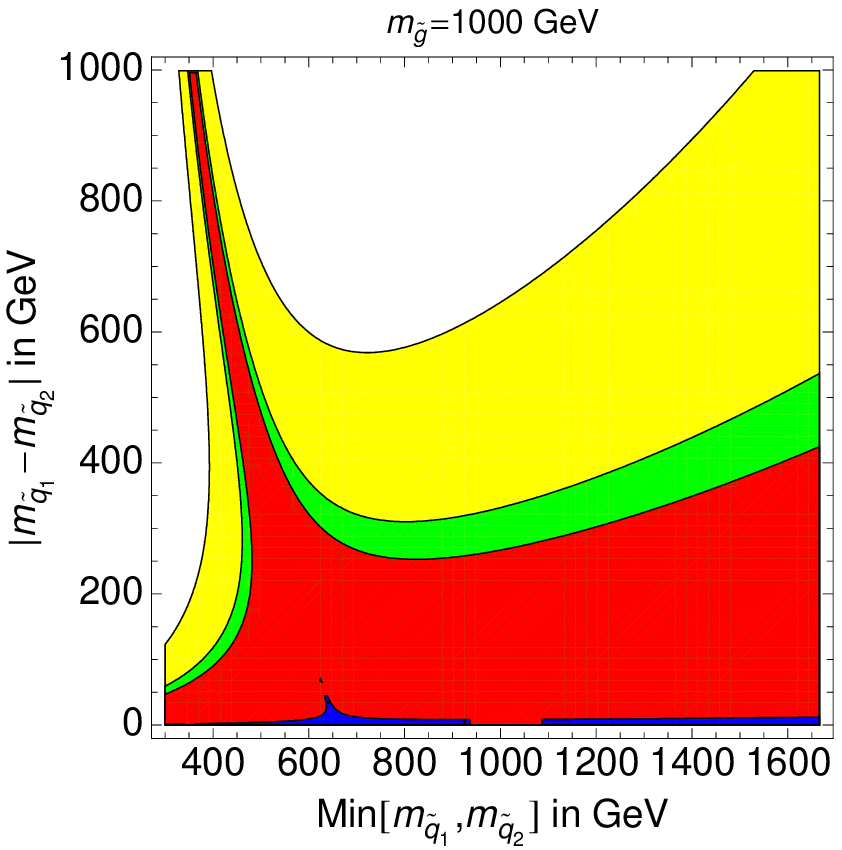}
\caption{\small Allowed mass splitting between the first two generations of left-handed squarks for different gluino masses. 
We assume the approximate GUT relation $M_2=(\alpha_2/ \alpha_s) m_{\tilde g}\cong 0.35$. Yellow (lightest) corresponds to the maximally allowed mass splitting assuming an intermediate alignment of $m^2_{\tilde{q}}$ with $Y_u^{\dagger}Y_u$ and $Y_d^{\dagger}Y_d$ \cite{Blum:2009sk}. The green (red) region is the allowed range assuming an diagonal up (down) squark mass matrix. The blue (darkest) area is the minimal region allowed for the mass splitting between the left-handed squarks, which corresponds to a scenario with equal diagonal entries in the down squark mass matrix but with an off-diagonal element carrying a maximal phase. Note that the allowed mass splittings are large enough to permit the decay of the heavier squark into the lighter one plus a W boson.}\label{mass-splitting}
\end{figure}

As stated before, SU(2) symmetry links a mass splitting in the up (down) sector to flavor-changing LL elements in the down (up) sector. So, if one assumes a ´´next-to minimal'' setup in which one mass matrix is diagonal, one has to specify if this is the up or the down squark mass matrix. If the down (up) squark mass matrix is diagonal, which implies that it is aligned to $Y_d^{\dagger}Y_d$ ($Y_u^{\dagger}Y_u$), one has contributions to \dd\ (\kk) mixing.
Assuming a diagonal up-squark (down-squark) mass matrix, the allowed regions compatible with \kk\ mixing (\dd\ mixing) are shown in Fig.~\ref{mass-splitting}. Note that there are large regions in parameter space with nondegenerate squark still allowed by \kk\ (\dd) mixing due to the cancellations between the different contributions shown in \fig{C1}. However, departing from an exact alignment with either $Y_u^{\dagger}Y_u$ or $Y_d^{\dagger}Y_d$ there are points in parameter space which allow for an even larger mass splitting \cite{Blum:2009sk} due to an additional off-diagonal element in the squark mass matrix. If this element is real one can choose an appropriate value which maximizes the allowed mass splitting \footnote{We thank Gilad Perez for bringing this to our attention.}. Nevertheless, this additional off-diagonal element now present in both sectors due to the SU(2) relation could also carry a phase additional to the CKM matrix. If this phase is maximal one obtains the minimally allowed range for the mass splitting due to the severe constraint from $\epsilon_K$. These minimally and maximally allowed regions for the mass splittings are also shown in Fig.~\ref{mass-splitting}.

This has interesting consequences both for LHC benchmark scenarios (which usually assume degenerate squarks for the first two generations) and for models with Abelian flavor symmetries (which predict non-degenerate squark masses for the first two generation) because \kk\ and \dd\ mixing cannot exclude non-degenerate squark masses of the first two generations. This allows for different decay chains of squarks. For example if the mass difference is larger than 80 GeV an additional W can be emitted leading to an additional jet or a charged lepton in the final state.
 
\section*{Acknowledgments}
I thank the organizers, especially Francesca Borzumati, for the invitation and the opportunity to participate in this very interesting workshop. I also thank Daniel Arnold for a careful proofreading of the article. A.C.~is supported by the Swiss National Science Foundation. The Albert Einstein Center for Fundamental
Physics is supported by the ``Innovations- und Kooperationsprojekt
C-13 of the Schweizerische Universit\"atskonferenz SUK/CRUS''.  

\bibliography{Trento}

\end{document}